\documentclass[aps,prc,twocolumn,floatfix,superscriptaddress,nofootinbib,preprintnumbers,longbibliography,amsmath,amsthm,amssymb]{revtex4-2}
\usepackage{graphicx}
\usepackage{amsfonts}
\usepackage{color}
\usepackage[colorlinks=true,linkcolor=blue,citecolor=blue]{hyperref}
\usepackage{dcolumn}
\usepackage{bm}
\usepackage{braket}

\newcommand{\br}{\boldsymbol{r}}

\newcommand{\ii}{\mathrm{i}}
\begin{document}
\allowdisplaybreaks[1]
\title{$\beta$-decay half-lives as an indicator of shape-phase transition in neutron-rich Zr isotopes with particle-vibration coupling effect}

\author{Kenichi Yoshida}
\email[E-mail: ]{kyoshida@rcnp.osaka-u.ac.jp}
\affiliation{Research Center for Nuclear Physics, Osaka University, 
Ibaraki, Osaka 567-0047 Japan}
\affiliation{RIKEN Nishina Center for Accelerator-Based Science, Wako, Saitama 351-0198, Japan}

\author{Yifei Niu} 
\email[E-mail: ]{niuyf@lzu.edu.cn}
\affiliation{MOE Frontiers Science Center for Rare Isotopes, Lanzhou University, Lanzhou 730000, China}
\affiliation{School of Nuclear Science and Technology, Lanzhou University, Lanzhou 730000, China}

\author{Futoshi Minato}
\email[E-mail: ]{minato.futoshi@phys.kyushu-u.ac.jp}
\affiliation{Department of Physics, Kyushu University, Fukuoka 819-0395, Japan}
\affiliation{RIKEN Nishina Center for Accelerator-Based Science, Wako, Saitama 351-0198, Japan}

\date{\today}
\begin{abstract}

\begin{description}
\item[Background]
$\beta$-decay half-life is sensitive to the shell structure near the Fermi levels.
Nuclear deformation thus impacts the $\beta$-decay properties. 
\item[Purpose]
A first-order shape-phase transition in neutron-rich Zr isotopes is predicted by some models.
We investigate the $\beta$-decay half-lives of neutron-rich nuclei around $^{110}$Zr, where the shape-phase transition is predicted to occur, to see if the $\beta$-decay half-life can be an indicator of the shape changes.
\item[Method] 
The proton--neutron quasiparticle random-phase approximation (RPA) is adopted to calculate the Gamow--Teller transitions.
In addition, we apply the quasiparticle phonon-vibrational coupling (PVC) to consider the phonon couplings.
\item[Results]
The spherical and oblate configurations 
give similar half-lives but shorter ones than the prolate configuration at the RPA level. 
The PVC effect further reduces the half-lives in general, but the effect is smaller for the deformed configuration than that for the spherical one. 
As a result, it makes the shape change from the oblate configuration to the spherical configuration visible.
Therefore, a sudden shortening of $\beta$-decay half-lives is always found at the nuclear shape changes.
\item[Conclusions]
$\beta$-decay half-life is an indicator of the shape-phase transition. 
The shape mixing and the roles of the triaxial deformation are subject to study in the future.
\end{description}
\end{abstract}
\maketitle
\section{Introduction}

The physics of exotic nuclei has been one of the major subjects in the field of nuclear science with the upgrading and constructing of the radioactive-ion (RI) beam accelerator facilities around the world.
Recent progresses in the development of the experimental technique of spectroscopic studies have unveiled the nuclear structure of exotic nuclei~\cite{nak17}, and 
it has attracted much interest in how the shape of a nucleus changes as a function of the number of neutrons and protons.

Empirical observables revealing the evolution of nuclear shape are 
the excitation energies of the $2^+_1$ and $4^+_1$ states and their ratio together with the $E2$ transition strengths. 
To explore the evolution of nuclear shells and deformations, 
the SEASTAR project~\cite{SEASTAR} has been undertaken at RIKEN RIBF, aiming at 
a systematic search for new $2^+$ energies in the wide range of neutron-rich nuclei.
Besides that, the two-neutron separation energies, the monopole transition strengths, and the isotope shifts also reflect the structural changes of neutron-rich nuclei~\cite{cam02}. 
Nuclear deformation also has a substantial impact on 
the high-frequency excitation modes, 
such as in the photoabsorption cross-sections~\cite{Yoshida:2010zu,Oishi:2015lph}.

The Zr isotopes with $A \simeq 100$ have been of 
theoretical and experimental interest in nuclear structure as a region of competition
between various coexisting prolate, oblate, and spherical nuclear shapes~\cite{cam02}. 
The first-order phase transition occurs uniquely in this region, while we usually see a gradual change of deformation with 
an increase in the neutron/proton number in other regions such as in the rare-earth nuclei.
The mean-field calculations rooted in nuclear density-functional theory (DFT)~\cite{lal99,sto03}, 
the macroscopic-microscopic calculation~\cite{mol95} as well as the recent shell model calculation~\cite{tog16} describe well the sudden change from the spherical to the deformed shape at $^{100}$Zr. 
The deformed region has been confirmed up to $^{110}$Zr by 
observing a low $E(2_1^+)$ value and the $R_{4/2}$ value being greater than three~\cite{pau17}.
Furthermore, the calculations~\cite{lal99,sto03,bla05,mol16} predict the shape transition from the deformed to the spherical configuration around $N=74$.

The $\beta$-decay half-life is one of the most experimentally accessible physical quantities for RI beam facilities and plays a decisive role in determining the time scale of the $r$-process nucleosynthesis~\cite{Nishimura:2012mh}. 
Observed short half-lives around $A \simeq 110$ region speed up the $r$-matter flow~\cite{nis11}.
There have been a considerable amount of works on the roles of the nuclear deformation on the Gamow--Teller (GT) strength distributions~\cite{Sarriguren:1999vj,Sarriguren:2005nk,Sarriguren:2010bi,Sarriguren:2014oba,Sarriguren:2015lga,Ha:2014fra,Ha:2016dqb,Ha:2017pek}. 
Then, it has been found that nuclear deformation plays an important role in the $\beta$-decay half-lives. 
These works are, however, based on the random-phase approximation (RPA) that considers coherent one-particle-one-hole excitations.
To understand nuclear excitations quantitatively, one sometimes needs to take into account the effect of beyond-RPA, namely higher-order configurations such as phonon-coupling effects and coherent two-particle-two-hole excitations.
Important roles of the beyond-RPA effect have been recognized also for the GT strengths, which give a crucial influence to $\beta$-decays.
The PVC effect is essential for reproducing the width of GT resonances \cite{Niu2014,Niu2016,Robin2019} and improving the $\beta$-decay half-lives~\cite{Niu2015,Niu2018,Litvinova2020}. 
We propose in this work the $\beta$-decay half-lives as an indicator of the shape-phase transition, which may give an impact on the $r$-process. 
We will demonstrate it within the Skyrme Hartree--Fock--Bogolibov (HFB) approach and proton--neutron quasiparticle-random-phase approximation (pnQRPA) under the condition of an axially-deformed shape.
We also discuss that one can confirm the shape-phase transition of neutron-rich Zr isotopes from the $\beta$-decay half-lives even in the presence of the phonon-coupling effect within the quasiparticle-vibration coupling (QPVC).

The paper is organized as follows.
In Sec.~\ref{method}, we briefly explain the models for evaluating the $\beta$-decay half-lives. 
In Sec.~\ref{results}, we show the results and discuss the roles of nuclear deformation and the effects of phonon coupling. 
Section~\ref{summary} summarizes the paper.

\section{Nuclear energy-density functional Method for $\beta$-decay properties}{\label{method}}
\subsection{Skyrme Hartree--Fock--Bogoliubov approach for nuclear deformation}{\label{HFB}}

In the framework of the nuclear energy-density functional (EDF) method we employ, 
the ground state of a mother nucleus is described by solving the 
HFB equation~\cite{dob84}.
The single-particle and pair Hamiltonians 
are given by the functional derivative of the EDF with respect to the particle density and the pair density, respectively. 
An explicit expression of the Hamiltonians is found in the Appendix of Ref.~\cite{kas21}. 
The average particle number is fixed at the desired value by adjusting the chemical potential. 
Assuming the system is axially symmetric, 
the HFB equation is block diagonalized 
according to the quantum number $\Omega$, the $z$-component of the angular momentum. 

\subsection{Proton--neutron quasiparticle random-phase approximation}{\label{QRPA}}

Since the details of the formalism can be found in Refs.~\cite{yos13,yos13e}, 
here we briefly recapitulate the basic equations relevant to the present study.
The excited states $| f \rangle$ in a daughter nucleus are described as 
one-phonon excitations built on the ground state $|\rm{RPA}\rangle$ of the mother nucleus as 
\begin{align}
| f\rangle &= \hat{\Gamma}^\dagger_{f} |{\rm RPA} \rangle, \\
\hat{\Gamma}^\dagger_{f} &= \sum_{\alpha \beta}\left\{
X_{\alpha \beta}^f \hat{a}^\dagger_{\alpha,{\rm n}}\hat{a}^\dagger_{\beta, {\rm p}}
-Y_{\alpha \beta}^f \hat{a}_{\beta,{\rm p}}\hat{a}_{\alpha,{\rm n}}\right\},
\end{align}
where $\hat{a}^\dagger_{\rm n} (\hat{a}^\dagger_{\rm p})$ and $\hat{a}_{\rm n} (\hat{a}_{\rm p})$ are 
the neutron (proton) quasiparticle (labeled by $\alpha$ and $\beta$) creation and annihilation operators 
that are defined in terms of the solutions of the HFB equation 
with the Bogoliubov transformation. 
The phonon states, the amplitudes $X^f, Y^f$ and the vibrational frequency $\omega_f$, 
are obtained in the pnQRPA with a cutoff at 60 MeV. 
The residual interactions entering into the pnQRPA equation 
are given by the EDF self-consistently 
except for the $J^2$ term: 
the $J^2$ term
in the EDF is neglected in the HFB calculation but included in the pnQRPA calculation. 

\subsection{Quasiparticle vibration coupling in spherical nuclei}{\label{PVC}}

The QPVC model includes correlations beyond the spherical pnQRPA model by taking into account the quasipaticle phonon coupling.
The self-energy of pnQRPA states is obtained by considering the coupling of doorway states consisting of a two-quasiparticle excitation coupled to a collective vibration. The properties of these collective vibrations, i.e., phonons $\lvert nL\rangle$, are  obtained by computing the QRPA response for states of natural parity 
$J^{\pi} =0^+$, $1^-$, $2^+$, $3^-$, $4^+$, $5^-$, and $6^+$, where those phonons with energy less than 20 MeV and absorbing a fraction of the non-energy weighted isoscalar or isovector sum rule (NEWSR) strength being larger than 5\% are taken into account the model space. 
The self-energy of the pnQRPA state $|f\rangle$ is given as 
\begin{eqnarray}
    \Sigma_f(E)&=&\sum_{\alpha \beta\alpha' \beta'} W^\downarrow_{\alpha\beta,\alpha' \beta'}(E) X_{\alpha \beta}^f 
    X_{\alpha' \beta'}^f \nonumber\\
    && + W^{\downarrow *}_{\alpha \beta,\alpha' \beta'}(-E)
    Y_{\alpha \beta}^f Y_{\alpha' \beta'}^f,
\end{eqnarray}
where $W^\downarrow_{\alpha \beta,\alpha' \beta'}(E)$ represents the spreading terms associated with the coupling of two-quasiparticle configurations with the doorway states, 
and the detailed expressions are given in Ref.~\cite{Niu2016}; $X^f$ and $Y^f$ are the forward and backward pnQRPA amplitudes, respectively, as defined in the last subsection but for the spherical case. To calculate the $\beta$-decay half-lives, 
we use Gaussian smearing to get the GT strength distribution, 
\begin{equation}
    S(E) = \sum_n \frac{1}{\sigma_n \sqrt{2\pi}} e^{-\frac{(E-E_n-\Delta E_n)^2}{2\sigma_n^2}} B_n,
\end{equation}
where $\sigma_n = (\frac{\Gamma_n}{2} + \eta) /\sqrt{2 {\rm ln} 2} $, with $\Delta E_n = {\rm Re} \Sigma_n(E)$ and $\Gamma_n=-2{\rm Im} \Sigma_n(E)$, and $B_n$ is the pnQRPA transition probability for state $|n\rangle$. $\eta$ is the averaging parameter in $W^\downarrow$ to avoid divergence, taken as 200 keV.  The details of the QPVC formulas can also be referred to Refs.~\cite{Niu2016,Niu2018}.

\subsection{Calculation of the $\beta$-decay half-lives}{\label{half_life}}

The $\beta$-decay half-life $T_{1/2}$ can be calculated with the 
Fermi's golden rule as~\cite{gov71},
\begin{align}
\dfrac{1}{T_{1/2}} &= \dfrac{\lambda_\beta}{\log 2} \notag \\
&=\dfrac{(g_A/g_V)^2_{\mathrm{eff}} }{D} \sum_{E_f^* < Q_\beta}f(Z, Q_\beta-E_f^*)
| \langle f | \hat{F} | {\rm RPA}\rangle  |^2, 
\label{beta_rate}
\end{align}
where $D=6147.0$ s and we set $(g_A/g_V)_{\mathrm{eff}}=1$ rather than its actual 
value of 1.26 to account for the quenching of the spin matrix in nuclei. 
The transition matrix element 
for the GT operator $\langle f|\hat{F}|{\rm RPA}\rangle$ is evaluated 
by the quasi-boson approximation as 
$\langle f|\hat{F}|{\rm RPA}\rangle \simeq 
\langle 0|[\hat{\Gamma}_f,\hat{F}]|0\rangle$, 
where $|0\rangle$ denotes the HFB ground state. 
The Fermi integral $f(Z,Q_\beta-E_f^*)$ in Eq.~(\ref{beta_rate}) 
including screening and finite-size effects is given by
\begin{equation}
f(Z,W_0) = \int_1^{W_0} p W(W_0 - W)^2 \lambda (Z,W) dW,
\end{equation}
with
\begin{equation}
\lambda(Z,W)=2(1+\gamma)(2pR)^{-2(1-\gamma)}e^{\pi \nu}
\left|\dfrac{\Gamma(\gamma + \ii\nu)}{\Gamma(2\gamma +1)}\right|^2,
\end{equation}
where $\gamma=\sqrt{1-(\alpha Z)^2}$, $\nu=\alpha ZW/p$, $\alpha$ is the fine structure 
constant, $R$ is the nuclear radius. 
$W$ is the total energy of $\beta$ particle, $W_0$ is the total energy available in $m_e c^2$ units, 
and $p=\sqrt{W^2-1}$ is the momentum in $m_e c$ units~\cite{gov71}. 
Here, the energy released in the transition from the ground state of the target nucleus 
to an excited state in the daughter nucleus is given approximately by~\cite{eng99}
\begin{equation}
Q_\beta -  E_f^* \simeq \lambda_\nu - \lambda_\pi + \Delta M_{n-H} - \omega_f.
\end{equation}

\subsection{EDF employed in the numerical calculations}{\label{EDF}}
We employ in the actual calculations a Skyrme-type 
EDF for the particle-hole channel. 
The SkM* functional~\cite{bartel82} is mainly used for the present investigation, and the SLy4 functional~\cite{Chabanat:1997un} is used to supplement the discussion. 
The pairing is considered by using the mixed-type 
contact interaction
\begin{equation}
V_{{\rm pp}}(\br,\br^\prime)=V_0\left[ 
1-\dfrac{1}{2}\dfrac{\rho(\br)}{\rho_0}
\right]\delta(\br-\br^\prime)
\label{eq:pair}
\end{equation}
with $V_0=-225$ MeV fm$^3$ and 
$-290$ MeV fm$^3$ for the SkM* and SLy4 functionals, respectively, and $\rho(\br)$ and $\rho_0$ being the isoscalar density and 
the saturation density $0.16$ fm$^{-3}$. 
The pairing strengths in the deformed HFB calculation here are determined to be consistent with the pairing energy in spherical HFB calculation for QPVC, where the pairing strengths are adjusted by the experimental pairing gap of $^{114}$Zr from three-point formulas.
In the pnQRPA calculations, 
we include the proton--neutron pairing interaction as 
Eq.~(\ref{eq:pair}) with the same strength.

\section{Results and discussion}{\label{results}}

\begin{figure}[t]
\begin{center}
\includegraphics[scale=0.35]{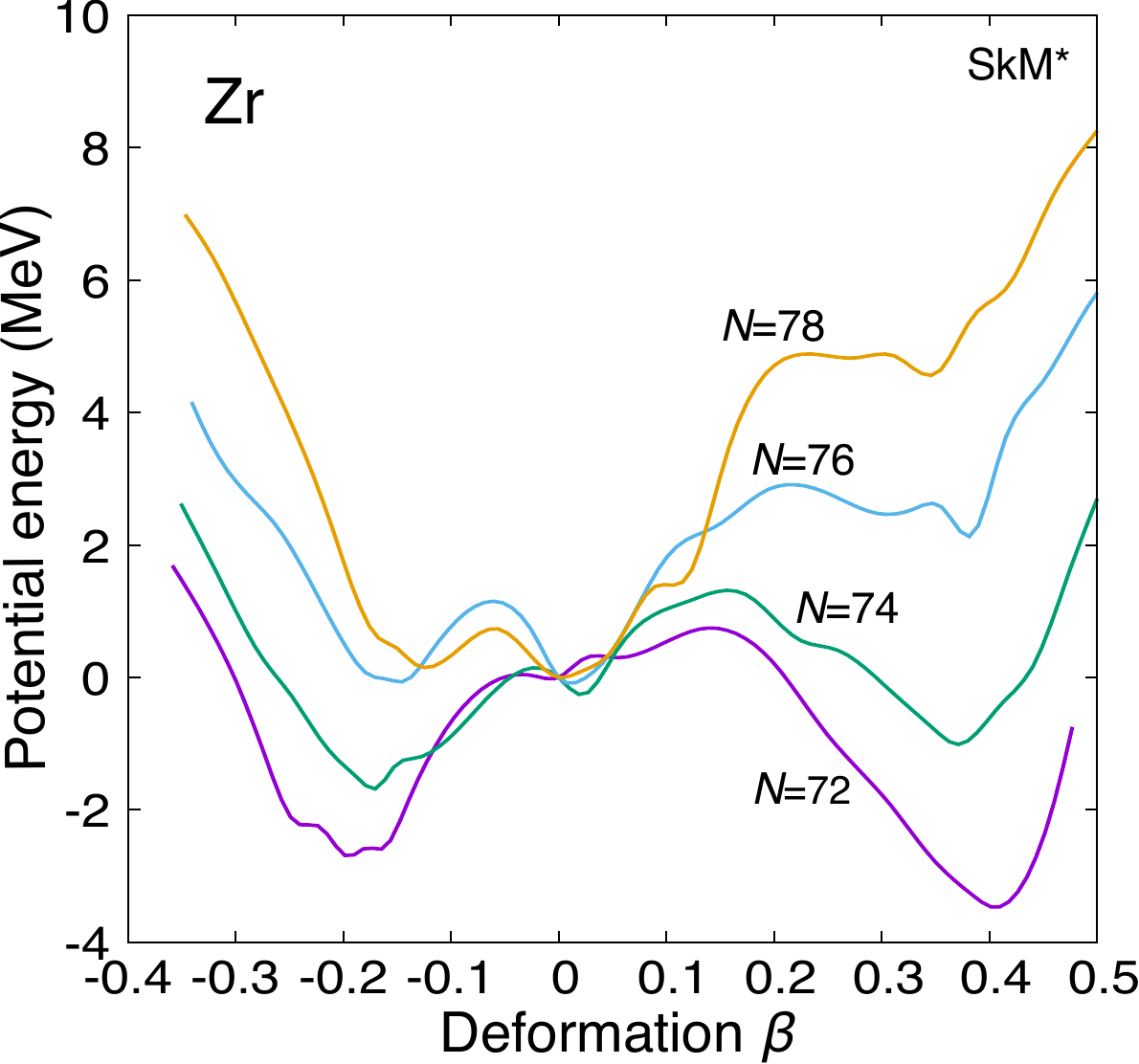}
\caption{\label{fig:Zr_PES} 
Potential energy surface (zero at the spherical configuration) of $^{112\text{--}118}$Zr obtained by employing the SkM* functional.}
\end{center}
\end{figure}

Figure~\ref{fig:Zr_PES} shows the potential energy surfaces of 
the neutron-rich Zr isotopes with $N=72 \text{--} 78$ calculated by 
using the SkM* functional, 
where the shape transition is predicted to occur with an increase in the neutron number~\cite{mol95,lal99,sto03,gen03,bla05}.
The prolate and oblate configurations compete in energy in $^{112, 114}$Zr, 
while the spherical and oblately deformed configurations compete in energy in $^{116,118}$Zr. 
We find a similar feature in the results obtained by employing the SLy4 functional, 
where the spherical and oblate configurations compete in energy.
A standard probe of the shape change from the prolate to oblate deformations is
a sign change of the spectroscopic quadrupole moment of the $2^+_1$ state. 
However, it is challenging to measure the spectroscopic quadrupole moment for these neutron-rich nuclei~\cite{gar21}.

\begin{figure}[t]
\begin{center}
\includegraphics[scale=0.4]{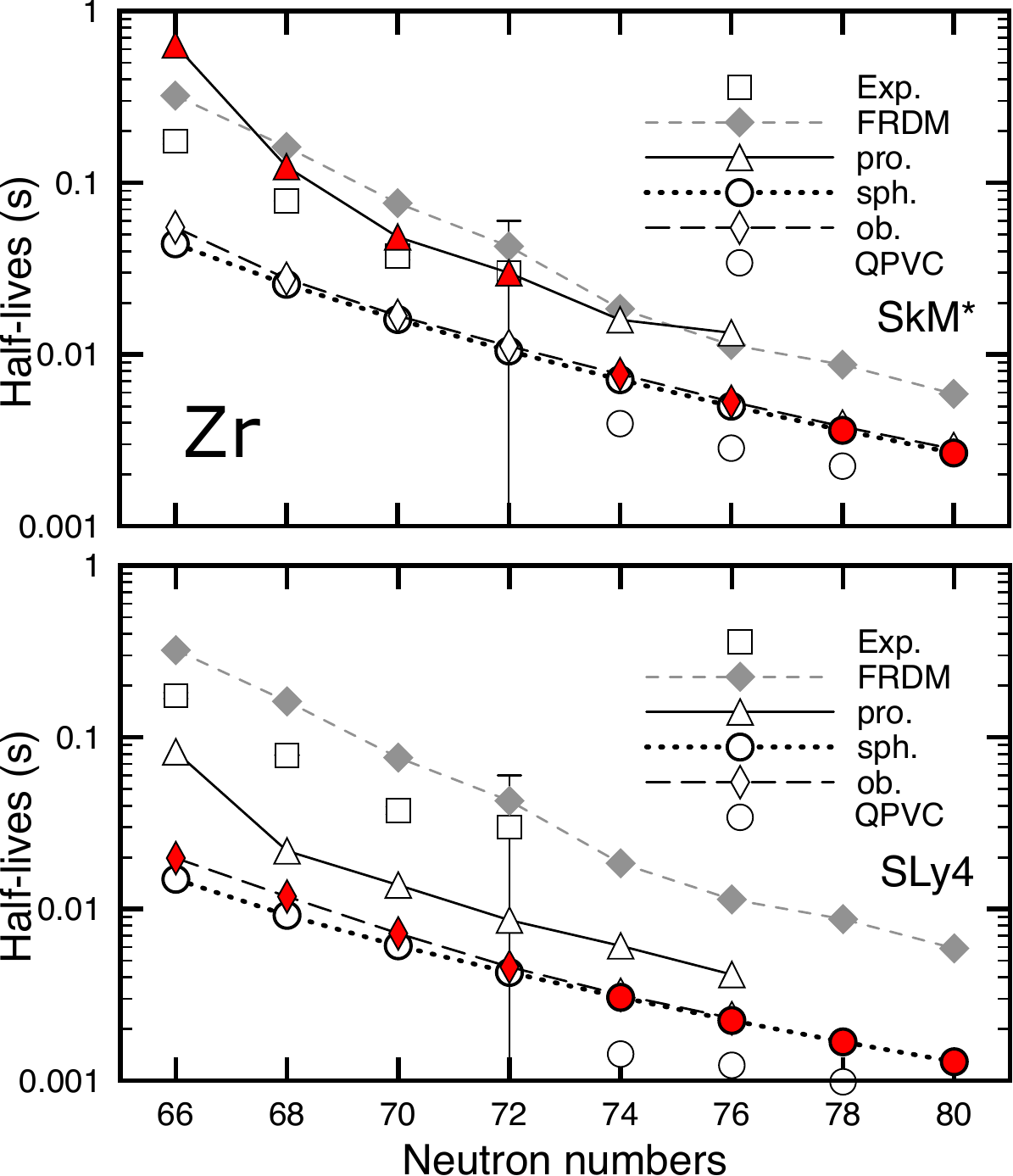}
\caption{\label{fig:Zr} 
$\beta$-decay half-lives of the Zr isotopes obtained by employing the SkM* (upper) and 
SLy4 (lower) functionals. 
Filled symbols indicate the lowest energy configuration. 
The calculated half-lives are compared with 
the experimental data~\cite{lor15} and the FRDM+QRPA calculation~\cite{mol03}. 
}
\end{center}
\end{figure}

\begin{figure}[t]
\begin{center}
\includegraphics[scale=0.35]{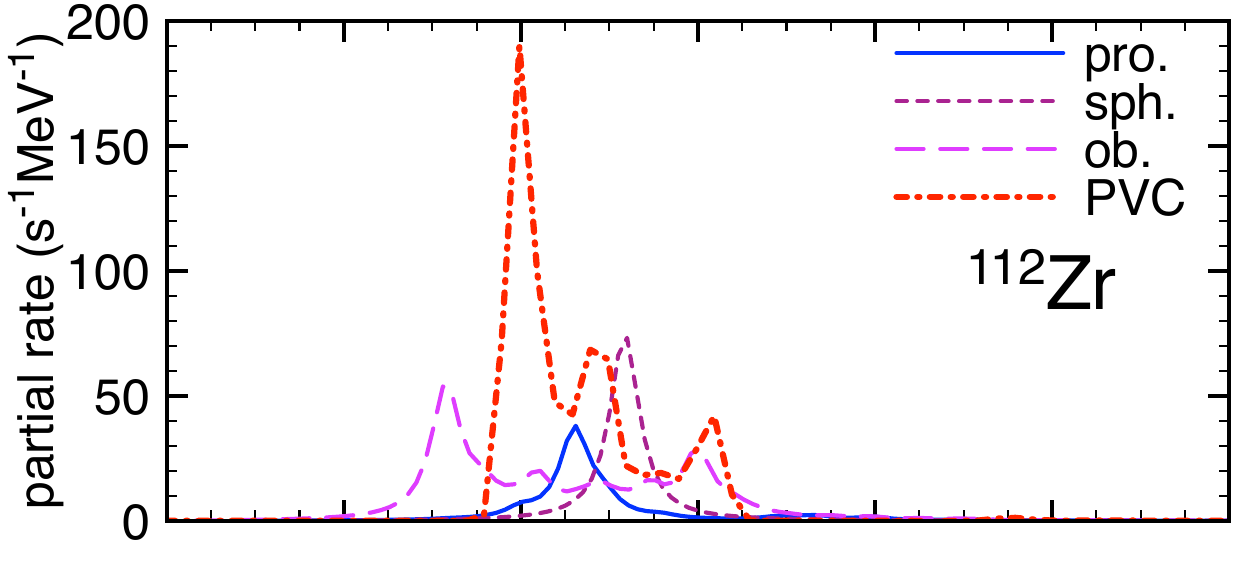}\\
\includegraphics[scale=0.35]{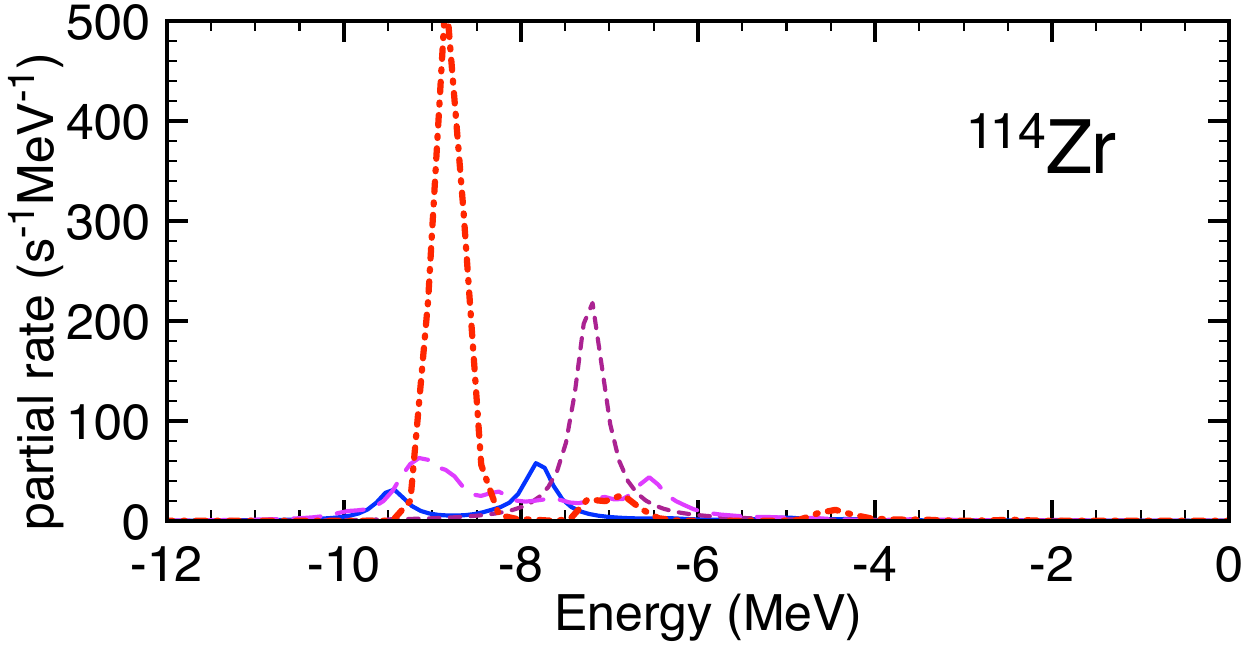}
\caption{\label{fig:Zr_GT} 
Distributions of the partial-decay rates in $^{112}$Zr (upper) and 
$^{114}$Zr (lower) as functions of the excitation energy with 
respect to the ground state of the mother nucleus. 
The RPA results for the prolate, oblate, and spherical configurations 
are shown together with the QPVC result for the spherical configuration. 
The RPA results are smeared by the Lorentzian function 
with a width of $\Gamma=0.4$ MeV.
}
\end{center}
\end{figure}

\begin{figure}[t]
\begin{center}
\includegraphics[scale=1.1]{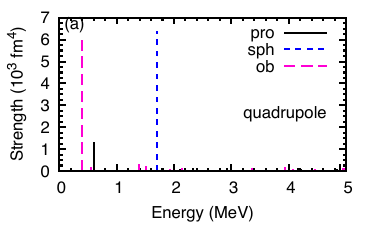}\\
\includegraphics[scale=1.1]{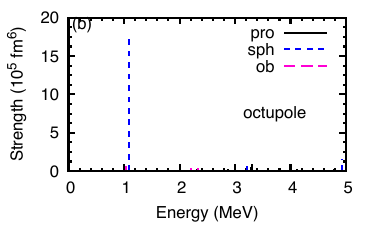}
\caption{\label{fig:114Zr} 
Distributions of (a) quadrupole and (b) octupole strengths in $^{114}$Zr calculated within the framework of Sect.~$\ref{HFB}$ and $\ref{QRPA}$.
}
\end{center}
\end{figure}

The $\beta$-decay half-life is a rather 
experimentally accessible quantity even for very neutron-rich nuclei,
and 
the calculated half-lives are shown in Fig.~\ref{fig:Zr}.
The observed half-lives up to $N=72$ are well reproduced by 
the calculation using the SkM* functional with the prolate configuration.
We see that the calculated half-lives calculated assuming the prolate shape shorten monotonically beyond $N=72$, 
wheres a sudden drop occurs at $N=74$ when the nuclear shape 
changes to the oblate deformation. 
In the case of SLy4, 
the results underestimate the measurements. 
However, we see the half-lives for the oblate configuration 
are shorter than for the prolate configuration as in case of SkM*.

We discuss the mechanism for the shortening of the 
half-lives due to the shape change 
from the prolate to oblate deformations. 
Figure~\ref{fig:Zr_GT} shows 
the distributions of the partial decay-rate 
associated with the GT transitions 
in $^{112,114}$Zr.
The GT states appear in low energies for the oblate configuration.
In $^{114}$Zr, the GT strengths for the oblate configuration are 
larger than those for the prolate configuration.
The GT strengths for the spherical configuration 
appear relatively higher in energy but are much
larger than those for the deformed configuration, 
leading to the half-lives being as short as 
for the oblate configuration.

We show in Figs.~\ref{fig:Zr} and \ref{fig:Zr_GT} the results considering the 
quasiparticle-phonon coupling for the spherical configuration denoted as QPVC.
Comparing with the results of half-lives for the spherical configuration, 
those for QPVC are shortened.
This is because the GT states couple with other phonon states, resulting in distributions to lower energies.
We will discuss the mechanism in more detail later on. 
It is considered that the PVC effect is weaker in deformed nuclei than 
in spherical nuclei because the quadrupole correlation is mostly 
described as a deformed mean field~\cite{nla.cat-vn1768774}.

Let us study the effect of PVC in the present case.
The low-lying phonon excitations are shown in Fig.~\ref{fig:114Zr}.
The first-excited quadrupole state appears at 1.7 MeV with a strength of $6.4 \times 10^3$ fm$^4$ for the spherical configuration,
while we see the $K=2$ state located around 0.6 MeV with a strength of $1.3 \times 10^3$ fm$^4$ for the prolate configuration. 
For the oblate configuration, the strength is $\sim 6.0 \times 10^3$ fm$^4$, which is 
much larger than that for the prolate configuration because these nuclei show 
softness against the triaxial deformation~\cite{Gogny_table}. 
We also show in the right panel of Fig.~\ref{fig:114Zr} the octupole excitations, the major states coupling with the GT states.
The calculated lowest-lying octupole state appears at 1.1 MeV both in the spherical 
and oblate configurations, while the strengths for the oblate and prolate configurations 
are more than one order of magnitude smaller than the case of the spherical configuration.
Since $^{114}\mathrm{Zr}$ has different behaviors for spherical and deformed shapes with respect to the quadrupole and octupole strengths, we need to investigate the interweaving roles of quadrupole and octupole phonons in PVC.
\begin{table}
\caption{Half-lives of $^{114}\mathrm{Zr}$ calculated by spherical RPA model (SkM*) and simplified PVC model (see text) with  only the first $2^+$ phonon (``PVC $2_1^+$ "), only the first $3^-$ phonon (``PVC $3_1^-$"), as well as both phonons (``PVC $2_1^+, 3_1^-$ "), taking the energies and transition strengths of these phonons at spherical and oblate deformation, respectively.}
\label{tab1}
\begin{tabular}{lcccc}
\hline\hline
Model & deformation & $T_{1/2}$ (ms) & deformation & $T_{1/2}$ (ms) \\
\hline
RPA & sph. & 6.4 & &\\ 
PVC $2_1^+$ & sph.  & 3.9 &  obl.  & 2.6 \\
PVC $3_1^-$& sph.  & 1.1 &  obl. & 5.4 \\
PVC $2_1^+, 3_1^-$ & sph. & 0.9 & obl. & 2.4 \\
\hline\hline
\end{tabular}
\end{table}

To distinguish the role of quadrupole and octupole phonons in PVC for $^{114}\mathrm{Zr}$, we consider a simple model.
In this simple model, we do not include the pairing correlations, as well as the momentum-dependent interactions in the PVC vertex calculation such that the transition densities of phonons could be used directly in the PVC vertex calculation~\cite{Niu2012}. 
With this approximation, we could estimate the PVC effect for deformed configurations by using the phonon energies of deformed configurations and rescaling the transition densities from the spherical configuration to the deformed one to adjust the transition strength. The corresponding results are shown in Tab. \ref{tab1}. 

From the spherical to oblate configurations, 
the lowest quadrupole-phonon energy is shifted downwards, and the strength increases, 
which should give a stronger PVC effect for the oblate configuration. 
This is confirmed in the simple model by including only the lowest quadrupole phonon in the PVC calculation, 
which gives 3.9 ms for the spherical configuration and 2.6 ms for the oblate configuration, compared with 6.4 ms in the RPA calculation. 
However, 
the lowest octupole phonon energy is nearly the same, but the strength is reduced by more than one order of magnitude, which would give a smaller PVC effect for the oblate configuration. 
This is confirmed by including only the lowest octupole phonon in the PVC calculation, 
which gives 3.9 ms for the spherical configuration and 5.4 ms for the oblate configuration. 
From the spherical to oblate configurations, the quadrupole and octupole phonons play different roles. 
Then we further include both phonons, and obtain 0.9 ms for the spherical configuration and 2.4 ms for the oblate configuration. 
It is clear to see that the PVC effect is much smaller for oblate configuration than that for the spherical configuration. 

As for the prolate configuration, the energy and strength of the lowest octupole phonon is similar to those in the oblate configuration (Fig.~\ref{fig:114Zr}(b)),
while the energy of prolate shape is higher than the oblate shape with the smaller strength for the lowest quadrupole phonon (Fig.~\ref{fig:114Zr}(a)). 
Thus, one can expect the PVC effect for the prolate configuration to be smaller than that for the oblate configuration. 
Therefore, after considering the PVC effect, the sudden change of half-lives from the prolate configuration to the spherical configuration remains, and the sudden change from the oblate to spherical configurations will also appear, which is not seen at the RPA level. 
For the SkM* functional, the shape change from the prolate to oblate configurations is observed at $N=74$, and the half-life is shortened already at the RPA level.
With the further inclusion of the PVC effect, the change in half-life will be more apparent.
For the SLy4 functional, the shape changes from oblate to spherical, and no significant shortening is observed in half-life at the RPA level, 
but with the further inclusion of the PVC effect, the sudden shortening of half-life will also manifest around $N=74$.

We have ignored the triaxial deformation in the present study. 
By looking at the PES in two dimensions of $\beta$ and $\gamma$ 
in Ref.~\cite{Gogny_table}, 
some nuclei are soft against the triaxial deformation. 
The beta-decay half-lives need to be investigated again after considering the traxial degree of freedom as well as the shape-mixing effect.

\section{Summary}{\label{summary}}

We have investigated the $\beta$-decay half-lives in the Zr isotopes with shape changes.
The GT strength distributions were evaluated in the proton--neutron QRPA and QPVC approaches.
The spherical and oblate configurations give similar half-lives, and the oblate configuration is shorter than the prolate one at the RPA level. 
The PVC effect further could reduce the half-lives; however, the effect would be smaller for a deformed configuration than that for a spherical one. 
When a sudden drop of half-lives around $N=74$ is observed experimentally, 
it is an indication of the shape transition.
However, the present model does not take into account the triaxial shape and the shape mixing.
Considering these effects remains a challenge in the future.

\begin{acknowledgments} 
This work was supported by the National Key Research and Development (R\&D) Program of China (Grant No. 2021YFA1601500), JSPS KAKENHI (Grants No. JP19K03824 and No. JP19K03872) and the JSPS/NRF/NSFC A3 Foresight Program ``Nuclear Physics in the 21st Century'', as well as the National Natural Science Foundation of China (Grant No. 12075104).
The numerical calculations were performed on the computing facilities  
at the Yukawa Institute for Theoretical Physics, Kyoto University, 
and at the Research Center for Nuclear Physics, Osaka University.  

\end{acknowledgments}

\end{document}